\documentclass[conference]{IEEEtran}
\IEEEoverridecommandlockouts
\usepackage{cite}
\usepackage{amsmath,amssymb,amsfonts}
\usepackage{algorithmic}
\usepackage{graphicx}
\usepackage{textcomp}
\usepackage{xcolor}
\usepackage{listings}
\usepackage{subcaption}
\usepackage[font=small]{caption}
\usepackage{multirow}
\usepackage{tabularx}
\usepackage[switch]{lineno}
\usepackage[flushleft]{threeparttable}
\usepackage{xspace}

\usepackage{amsmath}

\def\BibTeX{{\rm B\kern-.05em{\sc i\kern-.025em b}\kern-.08em
    T\kern-.1667em\lower.7ex\hbox{E}\kern-.125emX}}
    
\definecolor{ao}{rgb}{0.0, 0.5, 0.0}    
\newcommand{\tool}{\textsc{DeFP}\xspace}

\begin{document}

\title{Ranking Warnings of Static Analysis Tools Using Representation Learning\\
}

\author{\IEEEauthorblockN{
Kien-Tuan Ngo, 
Dinh-Truong Do, 
Thu-Trang Nguyen,
Hieu Dinh Vo
}

\IEEEauthorblockA{
Faculty of Information Technology, VNU University of Engineering and Technology, Vietnam \\
Email:\{tuanngokien, 17021090, trang.nguyen, hieuvd\}@vnu.edu.vn}
}

\maketitle

\thispagestyle{plain}
\pagestyle{plain} 

\begin{abstract}
Static analysis tools are frequently used to detect potential vulnerabilities in software systems. However, an inevitable problem of these tools is their large number of warnings with a high false positive rate, which consumes time and effort for investigating. In this paper, we present \tool, a novel method for ranking static analysis warnings. Based on the intuition that warnings which have similar contexts tend to have similar labels (true positive or false positive), \tool is built with two BiLSTM models to capture the patterns associated with the contexts of labeled warnings. After that, for a set of new warnings, \tool can calculate and rank them  according to their likelihoods to be true positives (i.e., actual vulnerabilities).
Our experimental results on a dataset of 10 real-world projects show that using \tool, by investigating only 60\% of the warnings, developers can find +90\% of actual vulnerabilities. Moreover, \tool improves the state-of-the-art approach 30\% in both Precision and Recall. 
\end{abstract}


\section{introduction}
\label{sec:introduction}

In order to guarantee the quality of software, many techniques such as code review, automatic static analysis, and testing, etc. have been applied during the software development life cycle. 
Especially, static analysis~\cite{ayewah2008using, nagappan2005static} plays an important role in detecting vulnerabilities at the early phases. 
Without executing programs, static analysis (SA) tools analyze the source code to identify the  violations of the
pre-defined rules and recommendations. 
These rules and recommendations are often defined by coding standards such as SEI CERT Coding Rule~\cite{CERT} or MISRA~\cite{MISRA}.

However, SA tools often report a large number of warnings (a.k.a. alarms)~\cite{beller2016analyzing}. 
In particular, a warning indicates the statement containing the potential vulnerability, the vulnerability type, and often the additional meta-information~\cite{flynn2018prioritizing}. 
In practice, developers have to manually inspect all the reported warnings and address them if necessary. However, due to the conservative over-approximation of program behaviors/properties of SA, many warnings 
are incorrectly reported by tools (i.e., false positive warnings).
%
Among the reported warnings, \textit{true positive warnings} or \textit{true positives} (TPs) are actual vulnerabilities, while \textit{false positive warnings} or \textit{false positives} (FPs) are the positions which indeed do not violate the checking rules.
Investigating FPs consumes time and effort but does not bring any benefit, therefore, the high FP rate reduces the productivity of developers~\cite{johnson2013don, koc2019empirical} and is bad for the usability of SA tools~\cite{johnson2013don, ruthruff2008predicting}. 
In consequence, it is necessary to reduce the number of FPs that developers need to verify.

In previous studies, sophisticated techniques such as model checking and symbolic execution have been applied to eliminate FPs~\cite{post2008reducing, li2013software, nguyen2019reducing}. Although these approaches obtain high precision in identifying FPs, they generally face non-scalability and time consuming issues because of their state space problems~\cite{muske2016survey}.

With the growth in size and complexity of the source code, recently machine learning (ML) techniques are leveraged to build models for discovering patterns associated with TPs/FPs. 
In general, detecting TPs/FPs among SA warnings can be considered as a standard binary classification problem. To build a SA waning classification model, there are two main methods for extracting features from the warnings, one uses a pre-defined set of features and the other encodes the features by ML models. Then, from the extracted features, the ML models need to predict whether the SA warning is a TP or FP.

In several existing studies~\cite{yuksel2014trust, flynn2018prioritizing, berman2019active}, the fixed sets of the hand-engineered features based on static code metrics and warning information have been derived from the source code and then fed to classifiers. 
%
The effectiveness of these approaches depends on the quality of the selected set of features. 
Moreover, the features in these approaches are manually defined by experts for certain kinds of warnings, therefore it is difficult to extend for handling different ones.  
%

Meanwhile, instead of using a fixed pre-defined set of features, Lee et al.~\cite{lee2019classifying} proposed a model to learn the lexical patterns of the statements around the warnings at the source code token level. 
They used \textit{Word2vec}~\cite{Mikolov2013EfficientEO} to embed code tokens into the vector form and then trained a Convolutional Neural Networks (CNN) classifier.  
Specially, by investigating their projects, they define the number of surrounding statements that need to be extracted to reflect the contexts of the warnings. These numbers of extracted statements are different for their six proposed checkers. 
However, it is challenging to apply this approach for different projects and/or different kinds of warnings since it requires expert knowledge and carefully manual investigation to decide how many statements in each function are enough to capture the contexts of the checking warnings.
Moreover, not every statement surrounding a warning is all related to its violation and equally important for TP/FP detection. Also, warning unrelated statements would negatively affect the performance of the ML models.  

In this paper, we propose \tool, a novel method to prioritize SA warnings.
Instead of classifying SA warnings, \tool predicts their likelihoods to be TPs and then ranks them with the top entries are more likely to be TPs (actual vulnerabilities), and the last entries are more likely to be FPs.
Indeed, ranking SA warnings rather than classifying them gives us three following benefits.
%

First, \textit{for developers, \tool helps shorten the cycle of developing and releasing products,  especially for critical systems.}
The reason is that critical systems are highly required to be safe, secure, and reliable. Therefore, any potential vulnerabilities (warnings) are all needed to be addressed. In other words, if a warning is eliminated due to being falsely classified as an FP, it would cause the system to be dangerously explored in the future. Instead of directly eliminating any warnings classified as FP, \tool 
ranks SA warnings according to their vulnerable likelihoods. 
With the pressure to release the high-quality software on time,
focusing on the top-ranked warnings first helps developers to find more actual vulnerabilities in a fixed duration. Then, they can spend time and effort on the low-ranked warnings later. 

Second, \textit{for SA tool builders, \tool suggests case studies that they can examine to improve the quality of their tools.}
Indeed, to better serve the market, SA tool builders need to frequently improve their tools by not only increasing the TP rate, but also decreasing the FP rate. Among a huge number of reported warnings in various projects, \tool suggests an effective order for investigation. Specially, SA tool builders can directly concentrate on addressing warnings which are more likely to be FPs, i.e., warnings at the last of the resulting lists, to find the patterns which tends to be incorrectly reported.

Third, \textit{for researchers, to build the datasets of real-world warnings, \tool helps the data collection process be more efficient.} In practice, in this field, it still lacks public datasets for evaluating approaches and researchers often have to manually investigate to label warnings.  This process is extremely time-consuming. From the ranked lists of \tool, researchers can effectively collect warnings by selectively labeling top-ranked and last-ranked warnings. 


Our key idea is based on the intuition that warnings which have similar contexts tend to have similar labels (TP or FP). For each warning, 
\tool exploits both syntax and semantics from the context of the warning and then
determines its likelihood to be TP.
In order to capture the context of a warning, \tool extracts all of the statements in the program which impact and are impacted by the  statement containing the warning (the reported statement). 
After that, to better represent the general patterns of warnings, identifiers and literals, which are specific for functions/files/projects and could make the ML models be biased by the training source code, are replaced by abstract names.
Next, the reported statements and their contexts are vectorized and used to train neural network models. One of 
the models represents the information specifically included in the reported statements, while
the other extracts critical information in the warning contexts. Then, the high-level features encoded from these models are utilized to estimate the likelihoods to be TPs of the corresponding warnings. Lastly, SA warnings are ranked according to their predicted scores.


To the best of our knowledge, it still lacks a public real-world dataset for widely evaluating approaches post-handling SA warnings.
In existing studies~\cite{yuksel2014trust, flynn2018prioritizing, berman2019active}, ML models are often trained and tested on synthetic datasets, such as Juliet~\cite{okun2013report} and SARD~\cite{SARD}. However, Chakraborty et al.~\cite{chakraborty2021deep} has demonstrated that these datasets are quite simple for estimating the performance of ML models on real-world data. 
Thus, to address the limitation of data shortage, we propose a dataset containing 6,620 warnings of 10 real-world projects.  
    
Our experiments show that
about 60\% of actual vulnerabilities are ranked by \tool in Top-20\% of warnings. Moreover, +90\% of actual vulnerabilities can be found by investigating only 60\% of the total warnings. Meanwhile, by using the state-of-the-art  approach~\cite{lee2019classifying}, with the same number of examined warnings, developers can find only 46\% and 82\% TPs.

In summary, our contributions in this paper are:
\begin{itemize}
    \item A novel approach to rank SA warnings, which does not require feature engineering and could be flexible to extend for different kinds of warnings. 
    \item A public dataset of 6,620 warnings collected from 10 real-world projects, which can be used as a benchmark for evaluating related work.
    \item An extensive experimental evaluation showing the performance of \tool over the state-of-the-art approach~\cite{lee2019classifying}.

\end{itemize}

\section{Motivating example and guiding principles}
\label{sec:approach_overview}

\subsection{Motivating Example}
\label{sec:motivating_example}

  

  

\begin{figure}
  \includegraphics[width=\columnwidth]{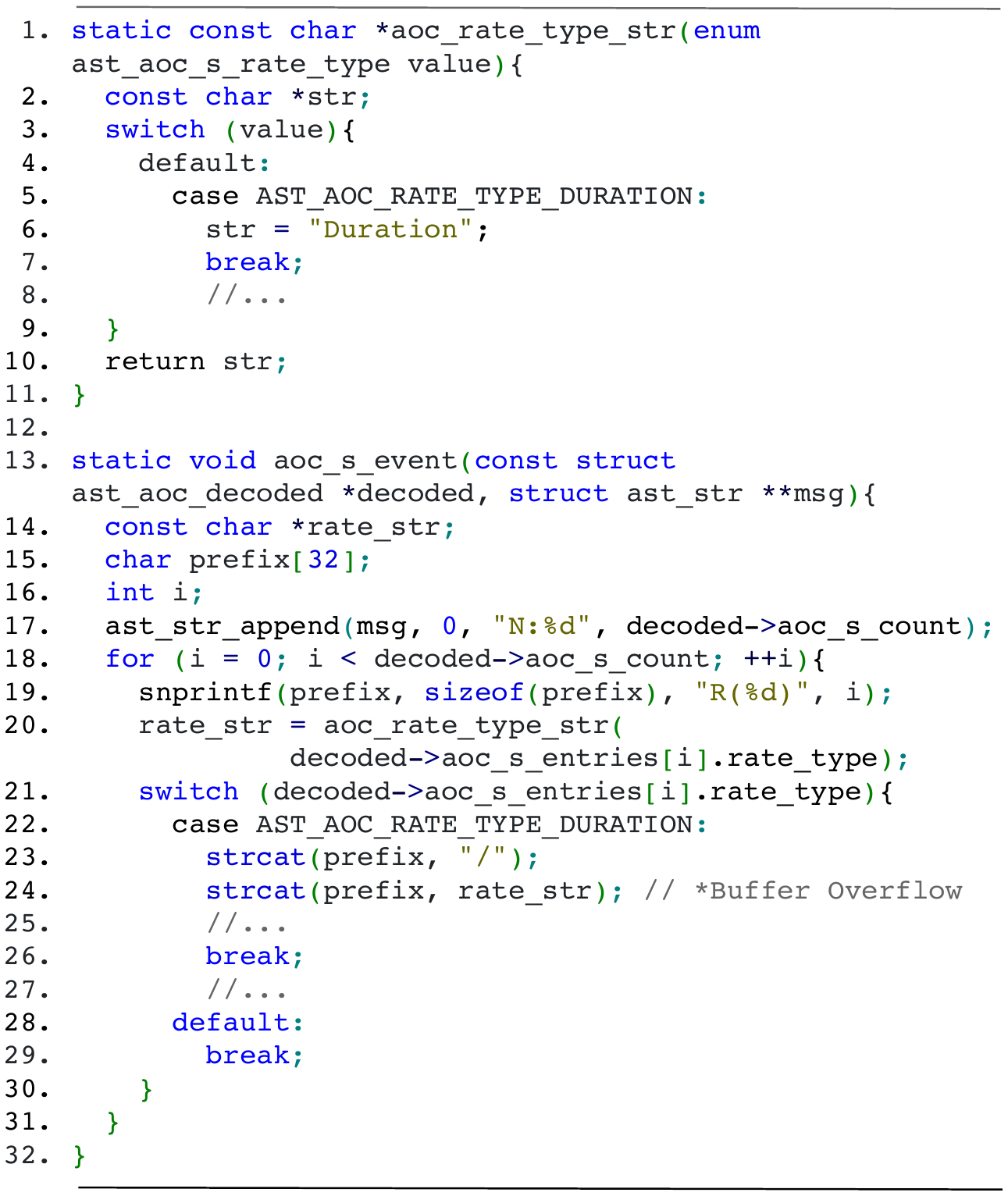}
  \caption{An FP reported by Flawfinder at line $24$}
\label{fig:example_code}
\end{figure}

Fig.~\ref{fig:example_code} shows a simplified version of function \texttt{aoc\_s\_event} in project \texttt{Asterisk}\footnote{https://github.com/asterisk/asterisk}, the complete version can be found on our website~\cite{website}. In this example, a warning related to \textit{Buffer Overflow} (BO) is reported at line $24$ by Flawfinder~\cite{FlawFinder}, a static analysis tool. The reason is that \texttt{strcat} appends the string pointed by \texttt{rate\_str} to the end of \texttt{prefix}. This could cause the size of the resulting string stored in \texttt{prefix} to be greater than $32$ which is the \texttt{prefix}'s size of allocated memory (line 15). 

However, via \textit{inter-procedure analysis}, we can conclude that this warning is an FP.
%
%
Specially, at line $19$ in Fig.~\ref{fig:example_code}, \texttt{prefix} is set to be \textit{R(i)} where \texttt{i} is the index of the loop, 
and the maximum size of \texttt{prefix} after this statement is $13$, in the case of $i = INT\_MAX$ (i.e., 2,147,483,647). 
At line $23$, \texttt{prefix} is appended a character (i.e., \texttt{"/"}), and then at line $24$ it is appended a string pointed by \texttt{rate\_str}, which has length $8$ (\texttt{rate\_str = "Duration"}, line $20$ and line $6$). As a result, after line $24$, the maximum length of \texttt{prefix} is $22$, which is still much smaller than $32$. 
Therefore, in order to determine whether a warning is a TP or FP,  
it need to conduct not only intra-procedure analysis but also inter-procedure analysis.
In other words, simply approximating the context of a warning by its surrounding statements or by its containing function could be ineffective.

\subsection{Guiding Principles}
In order to determine whether a SA warning is a TP or FP, analyzing only the reported statement is not enough. It requires investigating the context of the warning as well. For example, to conclude the warning at line $24$ in Fig.~\ref{fig:example_code} is an FP, not only that statement but also the related statements such as lines 18, 19, 20, etc. need to be examined. Therefore, to build an ML model which can
effectively predict the likelihoods to be TP/FP of the warnings, for each warning, we need to extract its appropriate context in the program.
From the extracted contexts, the model can capture patterns associated with the warnings. 
Also, 
statements unrelated to the warning,
which might cause noises and negatively affect the performance of the model, should be excluded from the warning context. 
In this paper, we propose the following principles for the problem of ranking SA warnings by representation learning.

\textbf{P1.} \textit{The warning contexts can be semantically captured by the statements in the program which can impact and be impacted by the reported statements}. 
In practice, to determine whether 
a warning indicates an actual violation of a specific vulnerability type or not, 
it is necessary to investigate all of the feasible execution paths containing the reported statement. In other words, we need to examine the control flows and data flows of the program which contain the reported statement. 
Besides, functions in a program do not work independently, they often execute with the invocations of the others. 
Thus, inter-procedural analysis is essential to effectively capture the contexts of the warnings.
Simply, all the statements in the functions/programs can be considered as the warning contexts, however, this method can cause unnecessarily large contexts and negatively affect the ML model's predictive performance. 
Also, not all of the statements in the functions/programs are actually relevant to the warnings. 
For example, the statement at line 17 in Fig.~\ref{fig:example_code} does not affect the decision about TP or FP of the warning at line 24. Including such irrelevant statements may cause the ML model to falsely learn the actual patterns associated with TPs/FPs. 
Therefore, inter-procedural slicing techniques~\cite{horwitz1990interprocedural} can be applied to effectively  extract the warning contexts by statements semantically related to the reported statements and eliminate the irrelevant statements. 


\textbf{P2.} \textit{The reported statements should be highlighted compared to the other statements in the warning contexts.} Intuitively, not all the statements in the program slices are equally important regarding the considering warnings. The reported statements are where the vulnerabilities are potentially explored, thus, they should be highlighted compared to the other statements in their contexts.
For example in Fig.~\ref{fig:example_code}, the statements at lines $19$, $23$, and $24$, which all modify the value of \texttt{prefix}, are all necessary for investigating the warning. However, according to the report of Flawfinder, the BO vulnerability is potentially explored at line $24$, which appends \texttt{prefix} by an unknown size string, i.e., a string is returned by another function. Intuitively, the statement at line $24$ should be emphasized compared to statements at lines $19$ and $23$.
Moreover, in practice, a program slice could contain several warnings, therefore to distinguish warnings, not only the program slices (i.e. context of the warnings) but also the reported statements need to be featurized.

\textbf{P3.} \textit{Identifiers should be abstracted.} 
The reason is that identifiers such as variables, function names, constants are project-specific (or even file-specific/function-specific) and considerably vary regarding developers' coding style. 
By learning such specific information, the ML model could not capture general patterns of the warnings.
Also, this could make the models simply learn the connections between specific identifiers and warning labels (TP/FP). Consequently, the models accurately predict the warnings of several training programs but their performance might decrease dramatically on the different programs.
Therefore, to build a general ML model which can work well and stably across programs, the identifiers should be abstracted into symbolic names, for example, \texttt{VAR1}, \texttt{FUNC1}, etc. 
Moreover, without abstraction, the number of identifiers could be virtually infinite, so ML model could have to deal with the vocabulary explosion problem.

\section{Static analysis warning ranking with representation learning}

\begin{figure*}
  \includegraphics[width=\textwidth]{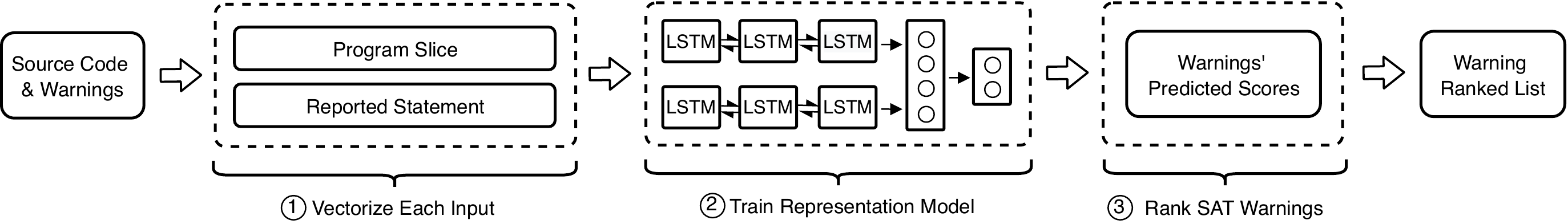}
  \caption{Our proposed approach for ranking SA warnings}
\label{fig:FPSAT_art}
\end{figure*}

Fig. \ref{fig:FPSAT_art} illustrates our SA warning ranking approach. 
Particularly, from the source code and the set of warnings of the analyzed program, we extract the reported statements and their program slices associated with warnings. For each warning, the reported statement and the corresponding program slice are converted into vectors and then fed to the BiLSTM models to predict its likelihood to be TP. After that, all of the warnings of the program are ranked according to their predicted scores.
\subsection{Program Slice Extraction}

In this work, to capture the context of each warning, we extract all the statements in the program which impact and are impacted by the corresponding reported statement. 
Specially, starting from the reported statement, we employ Joern~\cite{Joern} to conduct both backward and forward inter-procedural slicing in the program. 
%
%
For instance, the context of the warning at line 24 in Fig.~\ref{fig:example_code} is captured by the program slice shown in 
Fig.~\ref{fig:identifier_obfuscation_example}. 
%
Therefore, by this approach, not only a large number of irrelevant statements in the program are removed, 
but also the warning contexts are precisely captured via control/data dependencies relationship throughout the program. 

\subsection{Input Vectorization}

Program slices and reported statements are lexical source code tokens. Meanwhile, neural network models require their inputs to be formalized as numeric vectors. Therefore, we need to represent the model input data by a suitable data structure. In this section, we show three steps to represent the program slice and the reported statement of each warning: identifier abstraction, tokenization, and vectorization.

\subsubsection{Identifier Abstraction}

In general, programs often contain a huge number of identifiers, also their naming conventions and styles are diverse. It could cause difficulties for the ML models to capture the general patterns of the warnings~\cite{chakraborty2021deep}. 
Besides, ML models would simply learn characteristics of identifiers in certain projects, as well as simply map specific identifiers with corresponding warning labels. In order to avoid this problem, \tool abstracts all the identifiers before feeding them to the models. In particular, variables, function names, and constants in the extracted program slices are replaced by common symbolic names. For example, function name \texttt{aoc\_s\_event}  is replaced by \texttt{FUNC1}, the array \texttt{prefix} is replaced by \texttt{VAR5}, and its allocated size $32$ is replaced by \texttt{NUMBER\_LIT}. The details of our rules for abstracting identifiers can be found on our website~\cite{website}.

  

\begin{figure}
  \includegraphics[width=\columnwidth]{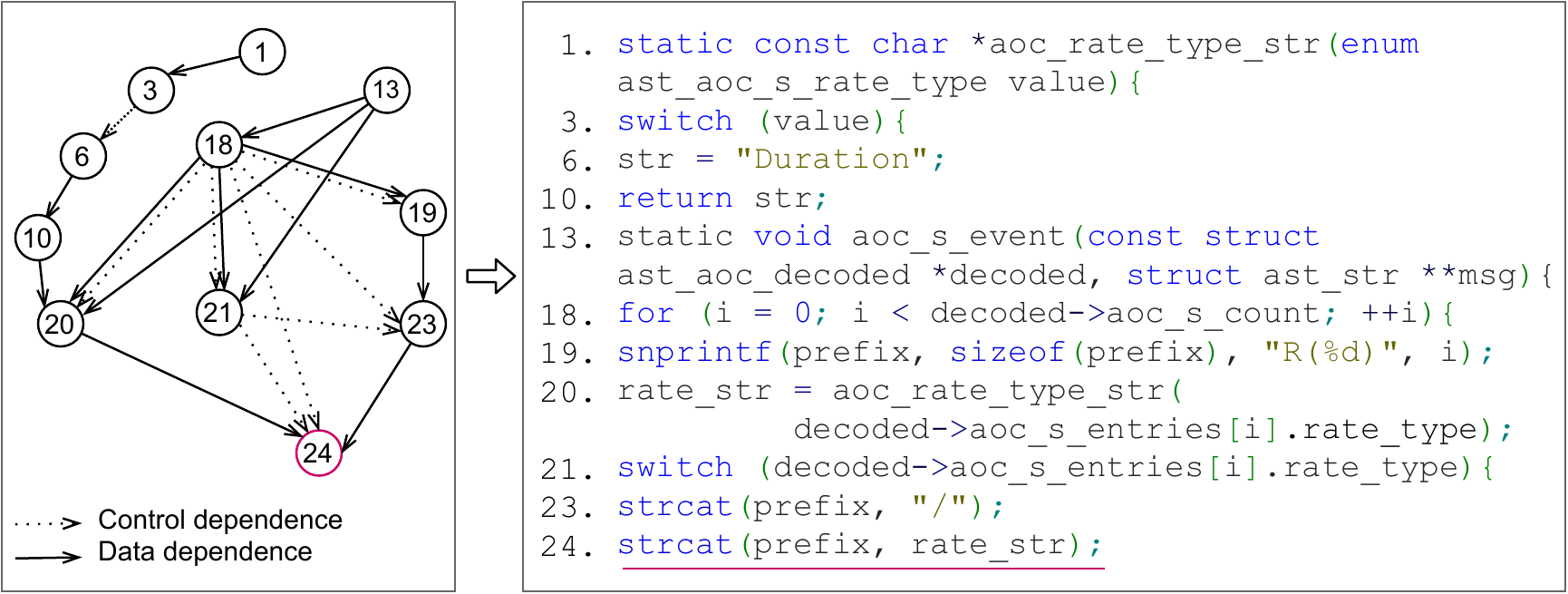}
  \caption{The program slice of the warning at line $24$ in Fig.~\ref{fig:example_code}}
\label{fig:identifier_obfuscation_example}
\end{figure}

\subsubsection{Tokenization}

To represent in numeric vectors for feeding to the neural network models, both the extracted program slices and the reported statements are tokenized into sequences of tokens. In this work, we use lexical analysis to break down each code statement into code tokens, which including identifiers, keywords, punctuation marks, and operators. 
For instance, the statement at line $24$ in Fig. \ref{fig:example_code}, \texttt{strcat(prefix, rate\_str);}, is abstracted as \texttt{strcat(VAR8, VAR11);} and then it is tokenized into seven separated tokens:  ``\texttt{strcat}'', ``\texttt{(}'', ``\texttt{VAR8}'', ``\texttt{,}'', ``\texttt{VAR11}'', ``\texttt{)}'' and ``\texttt{;}''.

\subsubsection{Padding and Truncation}
In practice, the number of sequence tokens in different slices could be significantly different. For example, in our experiment, the sequence lengths can vary from 5 to 9,566 code tokens. Therefore, to ensure that all the sequences have the same length, $L$, before inputting to the neural network, we use padding and truncation techniques. 
To achieve the best performance, the fixed length $L$ is carefully selected via multiple experiments. 
%

Particularly, for sequences having lengths smaller than $L$, we add one or more special tokens (\textit{$<$pad$>$}) at the end of these sequences. For the sequences whose lengths are greater than $L$, we truncate them to fit the fixed length.

In practice, the positions of the reported statements in their corresponding slices significantly vary. They can appear at the beginning of the slices or at the end of the slices. Thus, truncating from either the beginning or the end of the sequences could lead to the cases that the statements containing warnings are missed in the truncated sequences.
Therefore, in order to guarantee that the truncated sequences always contain the reported statements, we take these statements as the center for truncating.  
Specially, from the reported statements, we extend to both sides of the sequences. 
until reach the fixed length $L$.
Importantly, to capture the correct semantics of each code statement, we ensure that a statement will be fully included in the truncated sequences. 
It means that when only some tokens of a statement are included in the truncated sequences and the remaining is left due to the length limitation, we will replace all of the tokens of that statement in the truncated sequence by (\textit{$<$pad$>$}) token. 

\begin{figure*}[h!]
  \includegraphics[width=\textwidth]{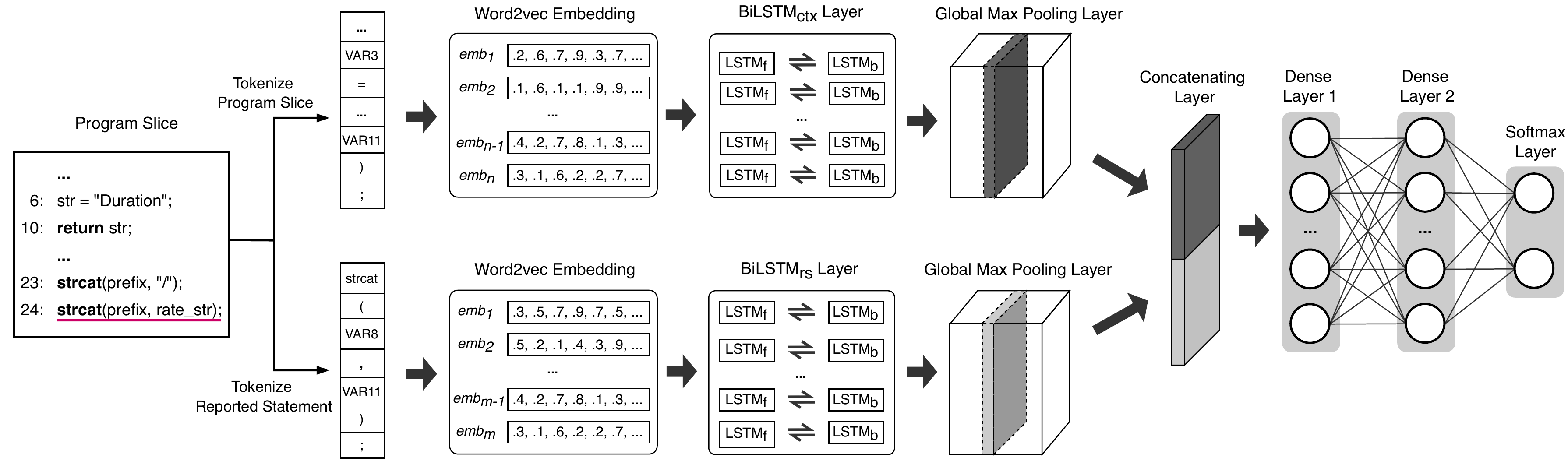}
  \caption{The proposed LSTM-based representation learning model for ranking SA warnings }
\label{fig:proposed_model}
\end{figure*}

\subsubsection{Vectorization}

In this step, the token sequences are embedded into numeric fixed-length vectors. 
In practice, besides the structural information of the sequences which need to be encoded, relationships between the tokens are also important. The reason is that code tokens have to appear together in a certain order to make the program grammatically and syntactically correct~\cite{lin2019software}.
For this purpose, in the vectorization step, we use \textit{Word2vec} model~\cite{Mikolov2013EfficientEO} with Continuous Bag-of-Words (CBOW) architecture.

\subsection{Representation Learning and Warning Ranking}

The \tool's model architecture is shown in Fig.~\ref{fig:proposed_model}.
Especially, to effectively learn contextual information that is crucial for revealing TP/FP code patterns, two Bidirectional Long Short-Term Memory networks (BiLSTM)~\cite{hochreiter1997long} are employed to train on the embedding vectors of the program slices and the reported statements. Afterwards, \tool extracts the meaningful characteristics of related warnings by concatenating the outputs from these BiLSTM models and feed into Fully Connected layers. In particular, we consider the final layer's output of the model as the likelihood to be TP/FP of each input warning. All these scores are finally gathered by \tool and ranked accordingly.

\subsubsection{Representation Learning}

In this work, the warning contexts (i.e. program slices) and the reported statements are encoded by two BiLSTM networks. In the case of program slices containing multiple statements distributed across functions, LSTM architecture might essentially apprehend the relationships between code tokens. 
%
Additionally, by utilizing a gated mechanism, LSTM can handle long-term dependencies and also focus on the most significant parts of the sequences. 

However, the information in LSTM is expressed one way through continuous time steps in sequential order.
Meanwhile, the occurrence of a code token is usually related to either the previous or the subsequent tokens, or even both.
%
Thus, additionally applying the bidirectional implementation of LSTM can assist the model to build dependencies in both forward and backward directions, which efficiently captures the general pattern of warnings.



\tool also employs the Global Max Pooling (GMP) layer to accumulate the output of each BiLSTM network. Especially, GMP layer computes maximum values over LSTM's time steps, which help to reduce output dimension and only keep the most important elements from LSTM cells. 
As a result, \tool has two GMP layers following each BiLSTM network, and then they are concatenated into a unified one to represent a whole feature map (Fig. \ref{fig:proposed_model}). 




\subsubsection{Warning Ranking}
After obtaining the learned representations of the warnings, \tool distinguishes their patterns by feeding them into three Fully Connected (Dense) layers behind.
Particularly, the final layer has only two hidden units activated by the Softmax function, which produces two scores whose total is 1.0. These two values correspond to the likelihoods of each warning to be TP and FP, respectively.

In the training phase, \tool's neural network enhances its predictions by finding the best hidden weights through estimating its errors. In other words, an objective function, cross-entropy, is applied to calculate the model's loss and update the weights towards minimizing this error value. Consequently, in the case of a TP warning, the model tends to converge its TP score towards 1.0 and its FP score closes to 0.0, and vice versa for an FP warning.
In the ranking phase, by inputting a list of warnings, \tool directly calculates their TP scores and sorts them in descending order.

\section{Empirical Methodology}
\label{sec:empirical_methodology}

In order to evaluate \tool, we seek to answer the following research questions:
\begin{itemize}
    \item \textbf{RQ1:} How accurate is \tool in ranking SA warnings? and how is it compared to the state-of-the-art approach~\cite{lee2019classifying}?
    \item \textbf{RQ2:} How does the extracted warning context affect \tool's performance? (P1)
    \item \textbf{RQ3:} How does the highlighting reported statement impact the performance of \tool? (P2)
    \item \textbf{RQ4:} How does the identifier abstraction component impact the performance of \tool? (P3)
   
\end{itemize}

\subsection{Dataset}

In order to train and evaluate an ML model ranking SA warnings, we need a set of warnings labeled to be TPs or FPs. Currently, most of the approaches are trained and evaluated by synthetic datasets such as Juliet~\cite{okun2013report} and SARD~\cite{SARD}. However, they only contain simple examples which are artificially created from known vulnerable patterns. Thus, the patterns which the ML models capture from these datasets could not reflect the real-world scenarios~\cite{chakraborty2021deep}. 
To evaluate our solution and the others on real-world data, we construct a dataset containing 6,620 warnings in 10 open-source projects~\cite{zhou2019devign, lin2019deep}. Table~\ref{tab:dataset} shows the overview of our dataset.

In these
10 real-world projects, functions are previously manually labeled as vulnerable and non-vulnerable
~\cite{zhou2019devign, lin2019deep}.
Then, our dataset is constructed by the following steps:
\begin{enumerate}
    \item \textit{Collecting warnings}: We pass the studied projects through three open-source SA tools, Flawfinder~\cite{FlawFinder}, CppCheck~\cite{CppCheck}, RATS~\cite{Rats} to collect a set of warnings.
    In practice, this set contains warnings in multiple kinds of vulnerabilities. However, we only collect the warnings related to \textit{Buffer Overflow (BO)} and \textit{Null Pointer Dereference (NPD)} 
    since for the other kinds, the number of reported warnings are too small for training and evaluating an ML model.
    \item \textit{Labeling warnings in the non-vulnerable functions}: Since, these functions are already marked as clean regarding BO and/or NPD, thus all the corresponding warnings
    in these functions are annotated as FPs.
    \item \textit{Labeling warnings in the vulnerable functions}: Although these functions are marked containing BO and/or NPD  vulnerabilities, we do not know exactly how many vulnerabilities each function contains and the positions of the vulnerabilities in the source code. Therefore, for each of the warnings in these functions, we manually investigate to label whether it is a TP or FP.
\end{enumerate}

\begin{table}[]
\centering
\caption{Overview of \tool's dataset}
\label{tab:dataset}
\resizebox{\columnwidth}{!}{%
\begin{threeparttable}
\renewcommand{\arraystretch}{1.1}
\begin{tabular}{|c|l||c|c|c||c|c|c|}
\hline
\multirow{3}{*}{\textbf{No.}} & \multirow{3}{*}{\textbf{Project}} & \multicolumn{3}{c||}{\multirow{2}{*}{\textbf{Buffer Overflow}}} & \multicolumn{3}{c|}{\textbf{Null Pointer}} \\[-0.5mm]
 &  &\multicolumn{3}{c||}{}  & \multicolumn{3}{c|}{\textbf{Dereference}} \\ \cline{3-8} 
 &  & \#W & \#TP & \#FP & \#W & \#TP & \#FP \\ \hline
1 & Asterisk & 2049 & 63 & 1986 & 133 & 0 & 133 \\ \hline
2 & FFmpeg & 1139 & 387 & 752 & 105 & 37 & 68 \\ \hline
3 & Qemu & 882 & 396 & 486 & 72 & 39 & 33 \\ \hline
4 & OpenSSL & 595 & 53 & 542 & 9 & 2 & 7 \\ \hline
5 & Xen & 388 & 15 & 373 & 23 & 6 & 17 \\ \hline
6 & VLC & 288 & 20 & 268 & 16 & 2 & 14 \\ \hline
7 & Httpd & 250 & 45 & 205 & 17 & 0 & 17 \\ \hline
8 & Pidgin & 250 & 13 & 237 & 242 & 0 & 242 \\ \hline
9 & LibPNG & 83 & 9 & 74 & 2 & 0 & 2 \\ \hline
10 & LibTIFF & 74 & 9 & 65 & 3 & 3 & 0 \\ \hline
 \# & \textbf{Total} & \textbf{5998} & \textbf{1010} & \textbf{4988} & \textbf{622} & \textbf{89} & \textbf{533} \\ \hline
\end{tabular}%
\begin{tablenotes}
    \item \#W, \#TP and \#FP are total warnings, true positives and false positives.
\end{tablenotes}
\end{threeparttable}}
\end{table}

\subsection{Evaluation Setup, Procedure, and Metrics}

\subsubsection{Experimental Setup}
We implemented neural network models using Keras together with TensorFlow backend (version 2.5.0). The tokenizer was built upon NLTK library (version 3.6.2) and \textit{Word2vec} embedding model was provided by the gensim package (version 3.6.0). All experiments were computed by a server running Ubuntu 18.04 with an NVIDIA Tesla P100 GPU.

We adopt cross-validation to train several neural networks and select the best parameter values corresponding to the effectiveness of predicting likelihoods to be TP warnings in the proposed dataset. Specially, for \tool, embedding size are set to 96, the maximum length of each slice and reported statement is fixed to 600 and 40, respectively, and they are learned by two BiLSTM networks which each has 256 hidden nodes. During the training phase, the dropout, batch size, and  number of epochs is set to 0.1, 64 and 60, respectively. Also, the minibatch stochastic gradient descent ADAMAX optimizer is selected with the learning rate of 0.002.

Besides, the data is sampled into stratified 5 folds, while 4 folds are picked for training and 1 remaining fold for testing (ratio of 8:2). We then run 5 different experiments on 5 pairs of training and testing data and aggregate average results for the final assessment of the corresponding experiment.

\begin{table*}[h!]
\centering
\caption{Performance of \tool and CNN model proposed by Lee et al.~\cite{lee2019classifying} in ranking SA warnings}
\label{tab:rq1}
\resizebox{\textwidth}{!}{
\renewcommand{\arraystretch}{1.17}
\begin{tabular}{|l||l||l||r|r||r|r||r|r||r|r||r|r|}
\hline
\multirow{3}{*}{\textbf{WN}} & \multirow{3}{*}{\textbf{Project}} & \multirow{3}{*}{\textbf{Method}} & \multicolumn{10}{c|}{\textbf{\# TP warnings found in top-k\% warnings}} \\ \cline{4-13} 
 &  &  & \multicolumn{2}{c||}{\textbf{Top-5\%}} & \multicolumn{2}{c||}{\textbf{Top-10\%}} & \multicolumn{2}{c||}{\textbf{Top-20\%}} & \multicolumn{2}{c||}{\textbf{Top-50\%}} & \multicolumn{2}{c|}{\textbf{Top-60\%}} \\ \cline{4-13} 
 &  &  & \multicolumn{1}{c|}{Precision} & \multicolumn{1}{c||}{Recall} & \multicolumn{1}{c|}{Precision} & \multicolumn{1}{c||}{Recall} & \multicolumn{1}{c|}{Precision} & \multicolumn{1}{c||}{Recall} & \multicolumn{1}{c|}{Precision} & \multicolumn{1}{c||}{Recall} & \multicolumn{1}{c|}{Precision} & \multicolumn{1}{c|}{Recall} \\ \hline
\multirow{8}{*}{BO} & \multirow{2}{*}{Qemu} & 
    CNN & 71.11\% & 8.09\% & 53.33\% & 12.13\% & 46.86\% & 20.72\% & 44.32\% & 49.25\% & 43.02\% & 57.57\%  \\
 &  & DeFP & 82.22\% & 9.34\% & 67.78\% & 15.40\% & 65.14\% & 28.78\% & 52.27\% & 58.08\% & 50.38\% & 67.43\% \\  \cline{2-13}
 & \multirow{2}{*}{FFmpeg} & 
    CNN & 30.91\% & 4.40\% & 31.30\% & 9.30\% & 33.24\% & 19.64\% & 32.46\% & 47.80\% & 33.04\% & 58.39\% \\
 &  & DeFP & 67.27\% & 9.56\% & 61.74\% & 18.34\% & 52.43\% & 31.00\% & 38.95\% & 57.37\% & 37.72\% & 66.66\% \\  \cline{2-13} 
 & \multirow{2}{*}{Asterisk} & 
    CNN & 11.00\% & 17.56\% & 8.78\% & 28.59\% & 7.56\% & 49.36\% & 4.49\% & 72.95\% & 3.82\% & 74.49\% \\
 &  & DeFP & 34.00\% & 53.97\% & 18.54\% & 60.26\% & 10.73\% & 70.00\% & 5.18\% & 84.10\% & 4.56\% & 88.97\% \\  \cline{2-13} 
 & \multirow{2}{*}{COMBINED} & 
    CNN & 43.00\% & 12.77\% & 39.67\% & 23.56\% & 34.25\% & 40.69\% & 25.40\% & 75.45\% & 23.46\% & 83.56\% \\
 &  & DeFP & 66.00\% & 19.60\% & 56.00\% & 33.27\% & 43.92\% & 52.18\% & 27.50\% & 81.68\% & 24.82\% & 88.42\% \\  \hline
\multirow{2}{*}{NPD} & \multirow{2}{*}{COMBINED} & 
     CNN & 63.33\% & 21.37\% & 43.33\% & 29.15\% & 38.40\% & 53.99\% & 21.29\% & 74.25\% & 19.62\% & 82.09\% \\
 &  & DeFP & 80.00\% & 26.93\% & 65.00\% & 43.66\% & 47.20\% & 66.14\% & 25.81\% & 89.74\% & 22.58\% & 94.25\% \\  \hline
\end{tabular}%
}
\end{table*}

\subsubsection{Empirical Procedure}\hfill

\textbf{RQ1.} We compare the performance of \tool and the CNN model proposed by Lee et al.~\cite{lee2019classifying} for ranking warnings in the proposed dataset.

\textbf{RQ2.} We study the impact of the warning contexts on the performance of \tool. Specially, we compare the performance of \tool in four scenarios of the warning contexts:
(1) the raw code of the program, (2) the program slices on control dependencies, (3) the program slices on data dependencies, and (4) the program slices on both control and data dependencies.

\textbf{RQ3.} We study the impact of highlighting the reported statements on \tool's performance. We compare the ranking results of \tool in two scenarios when the reported statements are and are not encoded for training the BiLSTM model.

\textbf{RQ4.} We study the impact of the identifier abstraction component by comparing the performance of \tool when the inputs are embedded with and without this component.

For evaluation, we have two experimental settings as widely adopted in related studies~\cite{dam2019lessons, lin2019deep, zhou2019devign}: within-project setting  and combined-project setting.
First, in \textit{within-project setting}, warnings from the same project are split into training and testing sets.
Second, in \textit{combined-project setting}, we collect the warnings from all 10 projects and then split them into training and testing sets.  
In practice, in several projects, the number of warnings is quite small for training and testing an ML model, and it could cause overfitting or underfitting problems. Thus, we only select three projects which have the largest number of BO warnings for evaluating RQ1 in the corresponding vulnerability type in the within-project setting.  RQ1 in the NPD vulnerability and the other research questions are only evaluated in the combined-project setting.

\subsubsection{Evaluation Metrics}

In order to evaluate \tool and compare its performance with the state-of-the-art approach, we applied Top-$k$\% Precision (P@K) and Top-$k$\% Recall (R@K). These two metrics are widely used in related studies~\cite{lin2019deep, nguyen2021comparison}, especially when the dataset is severely imbalanced. In this paper, P@K denotes the proportion of actual TP warnings in the Top-$k$\% of warnings ranked by the model, and R@K refers to the proportion of correctly predicted TP warnings in Top-$k$\% among the total actual TPs warnings. 
In particular, P@K and R@K are calculated using the following formulas, where $\{Actual \: TPs\}$ is the set of actual TP warnings, $\{Predicted \: TPs\}@K$ is the list of Top-$k$\% of warnings ranked first by the model.

$$P@K =  \frac{|\{Actual \: TPs\} \cap \{Predicted \: TPs\}@K|}{|\{Predicted \: TPs\}@K|}$$

$$R@K = \frac{|\{Actual \: TPs\} \cap \{Predicted \: TPs\}@K|}{|\{Actual \: TPs\}|} $$

\section{Experimental results}
\label{sec:experiment}

\subsection{Performance Comparison (RQ1)}

Table~\ref{tab:rq1} shows the performance of \tool and the CNN model proposed by Lee et al.~\cite{lee2019classifying} in Top-5\%--Top-60\% warnings of the ranked lists. Overall, \textit{\tool improves their model by nearly 30\% in  Precision and Recall for both BO and NPD warnings.}
For example, in FFmpeg, Qemu, and Asterisk with the Top-20\% of warnings returned by \tool, developers can find 23/79, 24/77, and 9/13 actual vulnerabilities. Meanwhile, by using the CNN model~\cite{lee2019classifying}, the corresponding figures are only 16/79, 15/77, and 6/13, respectively.
When the warnings of all the projects are combined, by investing 20\% of the warnings in the top of the ranked list, 105/202 actual BO vulnerabilities and 12/18 actual NPD vulnerabilities can be found by \tool, while these figures for the CNN model are only 82/202 and 10/18, respectively. 
Interestingly, 
with the results of \tool, developers can find +90\% of actual vulnerabilities by investigating only 60\% of the total warnings, which is 8\% better than the CNN model. 

Indeed, \textit{\tool obtains better results because it concentrates on statements which semantically describe the contexts of the warnings and \tool is not negatively affected by the statements which are unrelated to the warnings}. For example, the warning in Fig.~\ref{fig:example_code}, \tool captures its context by the statements which impact and are impacted by the reported statement at line 24 as shown in Fig~\ref{fig:identifier_obfuscation_example}. These statements are essential for semantically capturing the context of the warning because they show when and how the value of \texttt{prefix}, which is
reported to potentially overflow, is changed. 
Additionally, unlike the CNN model~\cite{lee2019classifying}, \tool ignores statements at lines 17 and 26, which do not play any role in reflecting the violation of the reported statement, although they are near it. Therefore, they could cause noises if they are encoded as the context of the warning. 
 
In addition, by inter-procedural analysis, \textit{\tool does not miss important information to validate the violation of the reported statements}.
Specially, there are statements, which are essential for validating the warnings, could be in multiple functions. 
For example in Fig~\ref{fig:example_code}, the statement at line 6 specifying the concatenated string to \texttt{prefix}
is extremely important to determine whether the BO vulnerability could occur at line 24 or not.
However, this statement is not in the same function with the reported statement, yet in another function \texttt{aoc\_rate\_type\_str}. 
This statement is captured by \tool, but it will be missed if only intra-procedural analysis is considered.



%
Interestingly, among the studied projects, \tool achieved the highest results in Qemu and the lowest results in Asterisk. Specially, for Top-20\% warnings of Qemu, \tool obtained 65.14\% in Precision, whereas this figure of Asterisk is only 10.73\%. The reason is that the models are impacted by the imbalance of the dataset. For instance, the numbers of TPs and FPs in Qemu are quite balanced, while they are greatly imbalanced in Asterisk. Moreover, Asterisk only contains 63 TPs, which is extremely small compared to its 1986 FPs.

\subsection{Impact of the Warning Context (RQ2)}


\begin{figure}
    \begin{subfigure}[b]{0.495\columnwidth}
         \includegraphics[width=\columnwidth]{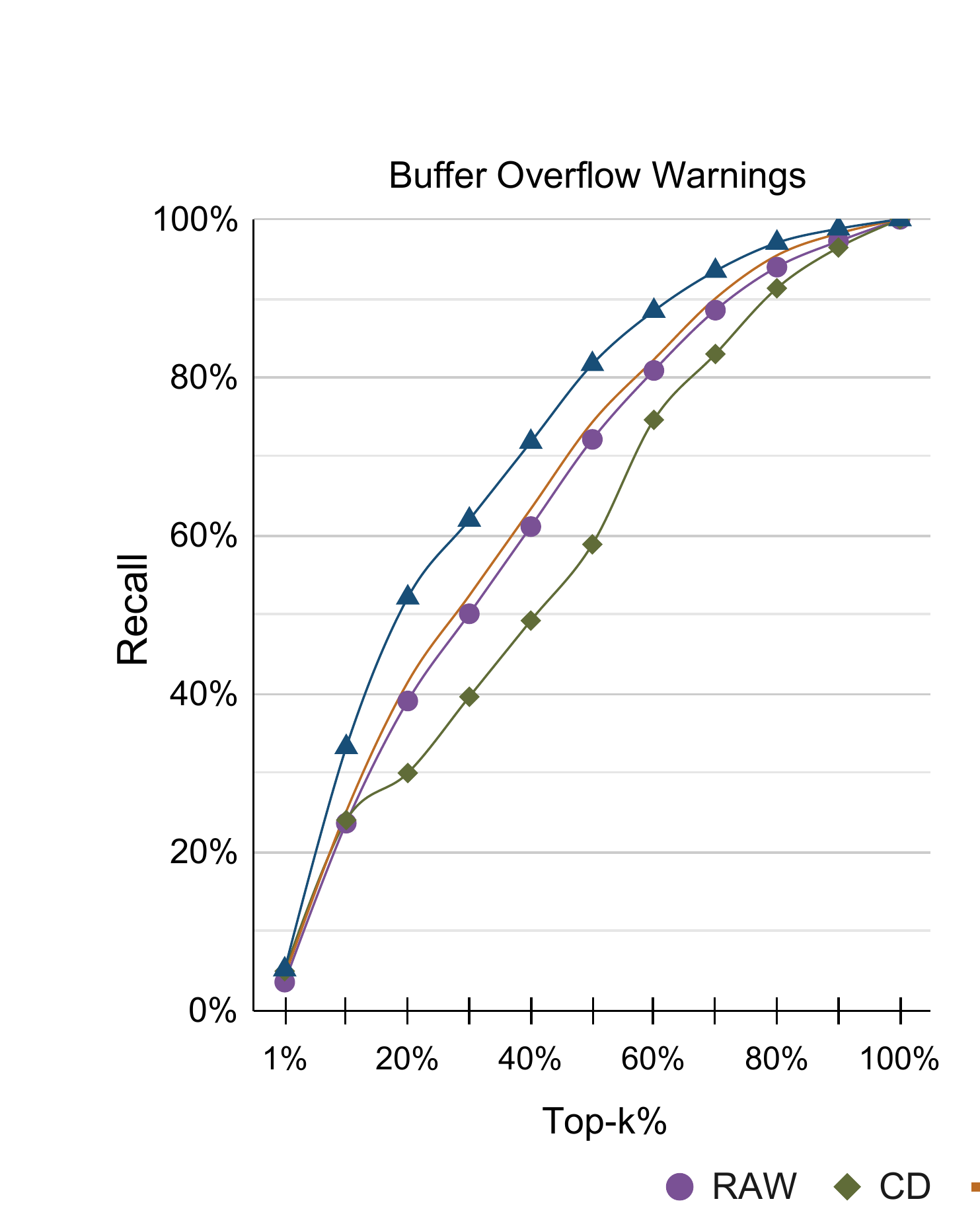}
     \end{subfigure}
     \begin{subfigure}[b]{0.495\columnwidth}
         \includegraphics[width=\columnwidth]{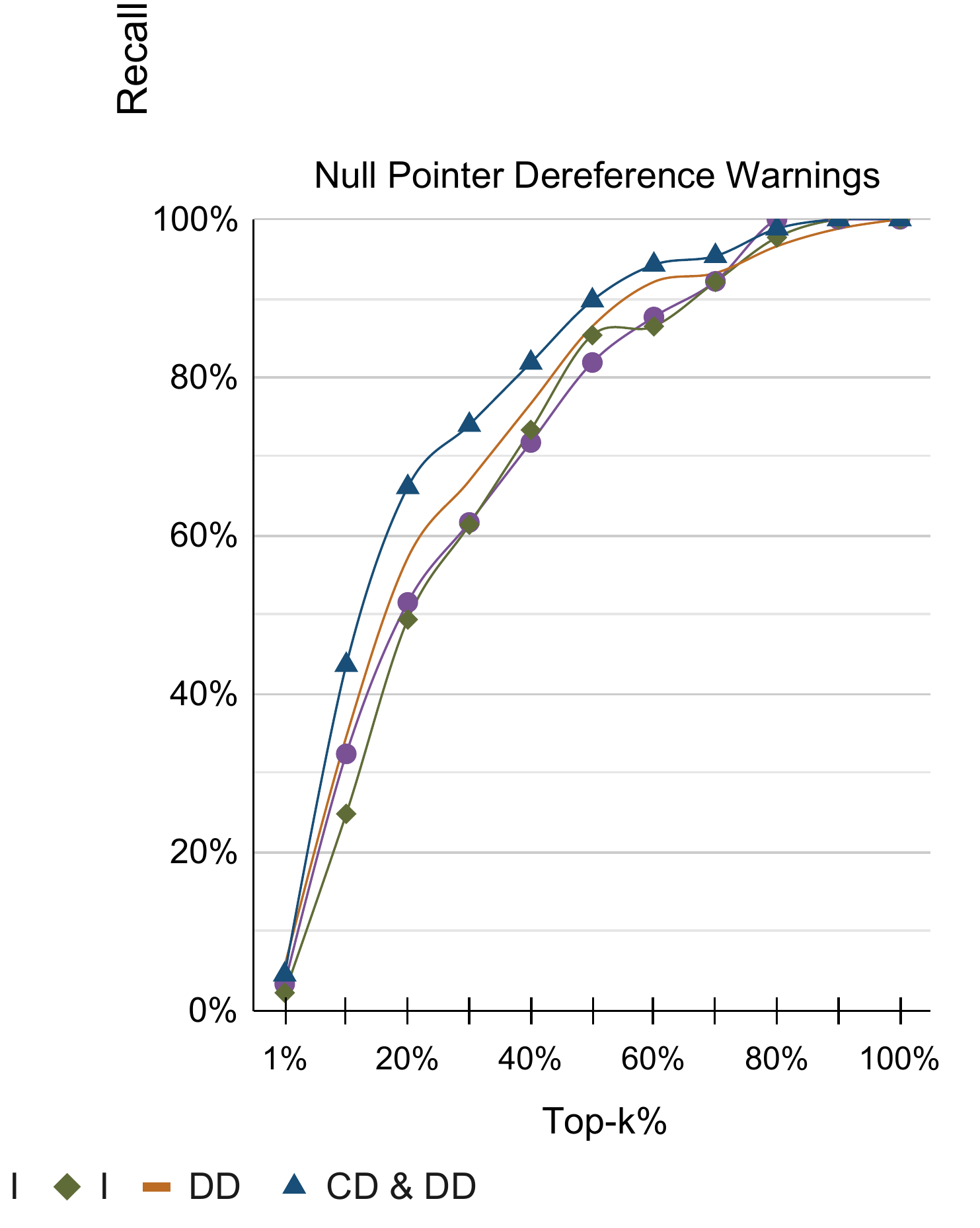}
     \end{subfigure}
  \caption{Impact of the extracted warning contexts on \tool's performance. RAW, CD, DD, and CD \&\& DD denote the warning contexts which are captured by raw source code, program slices on control dependencies, program slices on data dependencies, and program slices on both control and data dependencies, respectively.}
\label{fig:rq2_program_slice_impact}
\end{figure}

Fig.~\ref{fig:rq2_program_slice_impact} shows the performance of \tool when the contexts of the warnings are captured by different kinds of dependencies. 
\textit{\tool obtains the best performance when the warning contexts are captured by both control and data dependencies on the PDG. }
The reason is that, by using slicing techniques on both of these dependencies, unrelated statements are removed from the warnings' contexts and only related statements are encoded and fed to the BiLSTM models. Therefore, the models can better capture the patterns of warnings without being affected by noises caused by the statements which are semantically unrelated to the warnings.
In particular, when program slices are conducted on both control and data dependencies, the performance of \tool in two studied vulnerability types is 16\% and 19\% better than when the warning contexts captured by the raw code of the containing functions. Interestingly, by slicing on both control and data dependencies, \textit{the performance of \tool is significantly improved in the warnings which are ranked at the top of the lists.} Specially, compare to the raw code, for the Top-1\%--Top-50\% of the ranked warnings, \tool's performance with this kind of program slices is improved 42\% for NPD warnings and 34\% for BO warnings.
In other words, among Top-20\% of warnings (243 warnings for BO and 25 warnings for NPD) in the resulting lists, \tool correctly ranks 105/202 and 12/18 actual BO and NPD vulnerabilities. Meanwhile, when the warning context is captured by raw code of the whole functions, these figures are only 79/202 and 9/18, respectively. 

Importantly, \textit{for both BO and NPD vulnerabilities, 
program slices on only data dependencies capture the warning contexts better than the raw code of functions, however, the program slices on only control dependencies do worse.}
The reason is that
for these two kinds of vulnerabilities, the information about data dependencies, which illustrates how the values of the variables are propagated, is more informative for reasoning the warnings. 
For example in Fig.~\ref{fig:example_code}, to determine whether the warning (line 24) is an FP, it is essential to analyze the statements which have data-dependent on, such as lines 6, 15, 19, etc. 
Although the raw code may contain noises and unrelated statements, it still contains all of this information. 
However, this important information is missed in the program slices on control dependencies only.
Therefore, the performance of \tool with raw code is worse than the program slices on data dependencies, yet better than the program slices on control dependencies.
Specifically for Top-1\%--Top-60\% of warnings, compared to the raw code, \tool's results with program slices on the data dependencies is 7\% and 29\% better for BO and NPD.
Also, compared to the program slices on the control dependencies, these figures are 8\% and 46\%, respectively.

In practice, for different kinds of vulnerabilities, it could require control or data dependencies or both of these two kinds of information for validating the warnings. Therefore, to guarantee the best performance of \tool, program slices on both control and data dependencies should be leveraged to capture the warning contexts.

\subsection{Impact of the Reported Statements (RQ3)}

\begin{figure}
    \begin{subfigure}[b]{0.495\columnwidth}
         \includegraphics[width=\columnwidth]{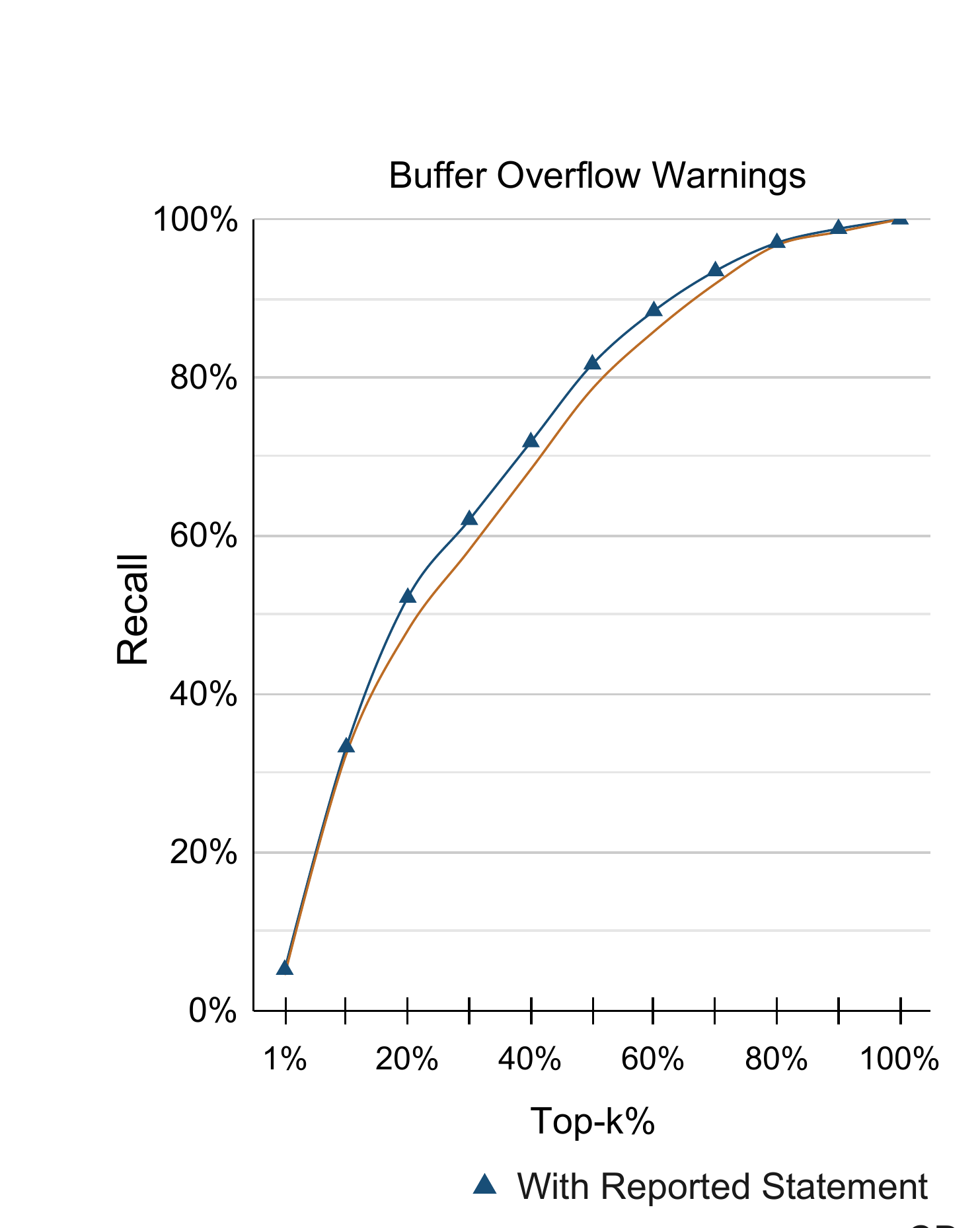}
     \end{subfigure}
     \begin{subfigure}[b]{0.495\columnwidth}
         \includegraphics[width=\columnwidth]{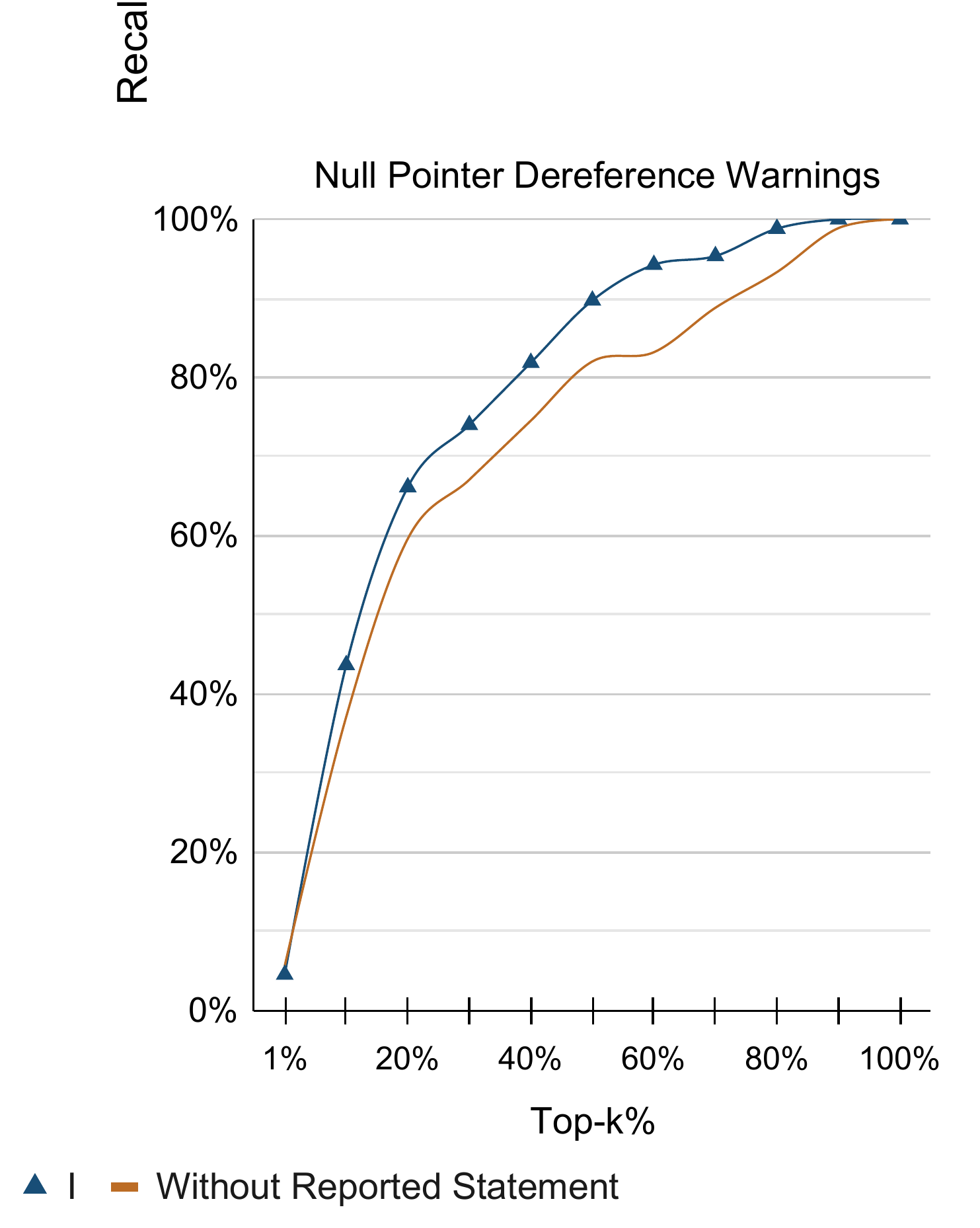}
     \end{subfigure}
  \caption{Impact of highlighting the reported statements on the performance of \tool}
\label{fig:rq3_reported_statement_impact}
\end{figure}
As seen in Fig~\ref{fig:rq3_reported_statement_impact}, \textit{the performance of \tool is slightly improved by 4\% and 7\% for BO and NPD vulnerabilities when the reported statements are highlighted by being encoded as an input of the BiLSTM model}. 
For example, in Top-20\% of ranked warnings, by encoding the reported statements, developers can find 7 more actual BO vulnerabilities and 1 more actual NPD vulnerabilities. 
More details about the performance of \tool on P@K can be found on our website~\cite{website}.

Indeed, highlighting the reported statements can help the neural network model not only capture the patterns associated with the warning contexts, but also explicitly emphasize the positions of warnings.
Consequently, this would be considerably helpful when several warnings having similar contexts but labeled (TP and FP) differently. 
However, our dataset is built from the set functions which are already classified as vulnerable or non-vulnerable.  
Thus, most of the warnings in a vulnerable function tend to have the same TP labels. Also, all of the warnings in a non-vulnerable function are labeled as FPs.
That is the reason why the \tool's performance is just slightly improved when the reported statements are encoded as an input of the representation model.

\subsection{Impact of the Identifier Abstraction Component (RQ4)}

\begin{figure}
    \begin{subfigure}[b]{0.495\columnwidth}
         \includegraphics[width=\columnwidth]{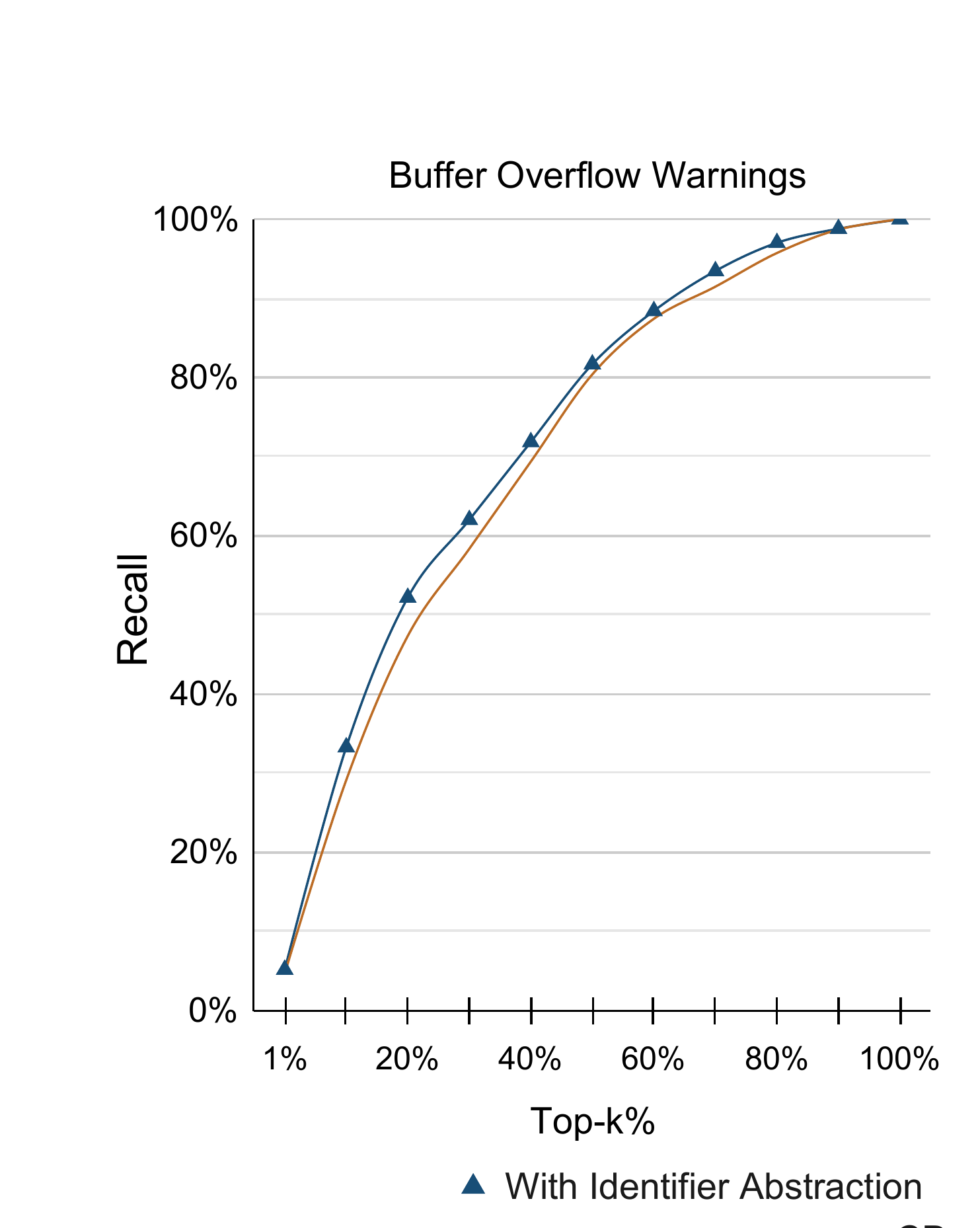}
     \end{subfigure}
     \begin{subfigure}[b]{0.495\columnwidth}
         \includegraphics[width=\columnwidth]{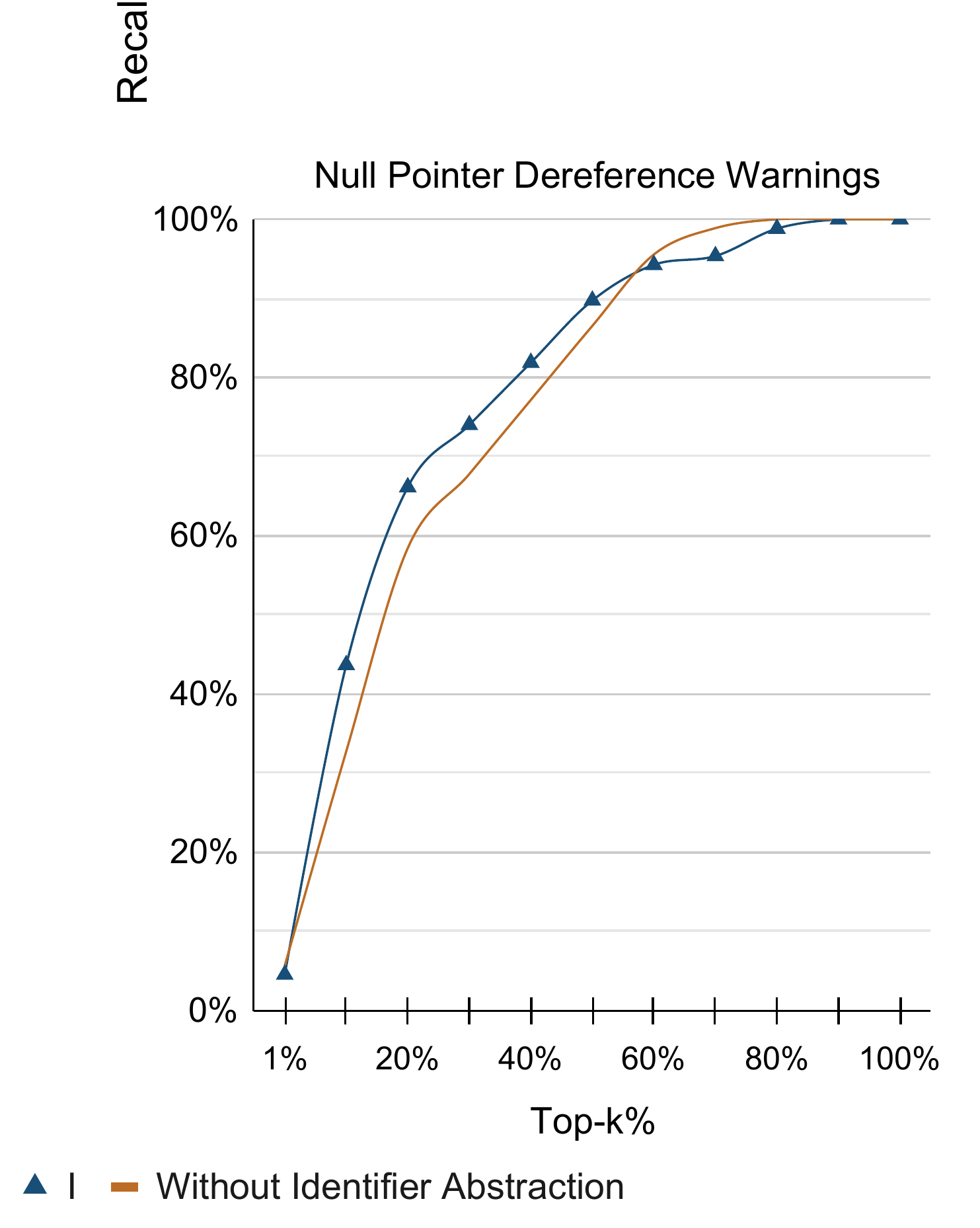}
     \end{subfigure}
  \caption{Impact of identifier abstraction on \tool's performance}
\label{fig:rq4_identifier_abstraction_impact}
\end{figure}
Fig.~\ref{fig:rq4_identifier_abstraction_impact} shows that \textit{by abstracting identifiers, \tool can capture the general patterns associated with the warnings better.}
Specially, with identifier abstraction, \tool achieves about 7\% and 12\% better in two kinds of studied vulnerabilities for Top-1\%-Top-60\%.
For instance, in Top-20\% of warnings, \tool can find 105 actual BO vulnerabilities and 12 actual NPD vulnerabilities, which is about 52\%  and 66\% of their total actual vulnerabilities. Meanwhile, without identifier abstraction, these numbers are only 96 and 11 vulnerabilities, respectively.
More details about the performance of \tool on P@K can be found on our website~\cite{website}.

Moreover, identifier abstraction decreases \textit{Word2vec} vocabulary size on BO and NPD datasets from 37,170 to 512 and 4,094 to 259 tokens, respectively. This helps the model deal with the vocabulary explosion problem, better generalize rare identifiers, and avoid out-of-vocabulary. As well,  \textit{Word2vec} might also beneficially reduce vector dimension to represent each code token, thus improve memory usage and shorten the training/prediction time.

\section{Threats to Validity}
There are three main threats to validity in this paper, they are external validity, internal validity, and construct validity, respectively, which are illustrated as follows.

\subsubsection{External validity} Our dataset contains 10 open-source programs and warnings in only two vulnerability types. Therefore, our results may not be general for all software projects and other kinds of vulnerabilities. To reduce the threat, we chose the programs which are widely used in the related work~\cite{zhou2019devign, lin2019deep} and the top most popular vulnerabilities~\cite{li2018vuldeepecker}.
Also, we plan to collect more data for the future work.

\subsubsection{Internal validity} For this paper, the internal validity mainly lies in the data used for learning process. We manually  labeled for the warnings based on the labels of the functions, which are assigned by Zhou et al.~\cite{zhou2019devign} and Lin et al.~\cite{lin2019deep}. The threat may come from their incorrect labels at function level or our misleading labels at warning level. To minimize this threat, we carefully investigate to label the warnings.

\subsubsection{Construct validity} In this study, we adopt P@K and R@K for evaluating the performance of the ranking models. However, with the problem of handling SA warnings, evaluation in terms of other metrics may also require in practice. We will conduct experiments using more evaluation measures in our future work.

\section{Related Work}
\label{sec:related_work}

There are various approaches have been applied to detect source code vulnerabilities in the early phases of the software development process. 
Specially, using SA tools is an automatic and simple way to detect various kinds of vulnerabilities without executing the programs~\cite{ayewah2008using, nagappan2005static}. Baca et al.~\cite{baca2009static} has demonstrated that SA tools is better than average developers in detecting warnings, especially the security ones. However, the generated warnings of SA tools often contain a high number of FP rate~\cite{johnson2013don, koc2019empirical}. Therefore, developers still need to waste a lot of time and effort for investing such FP warnings.

In order to improve the precision of SA tools, sophisticated program verification techniques such as model checking, symbolic execution, or deductive verification, etc., have been applied to reduce the number of FPs~\cite{muske2013efficient, muske2015efficient, nguyen2019reducing, post2008reducing, li2013software}. For instance, Muske et al.~\cite{muske2013efficient, muske2015efficient} uses model checking to eliminate FPs.  Specially, for each warning, they generate appropriate assertions and then use model checking to verify whether those assertions hold. Nguyen et al.~\cite{nguyen2019reducing} also generate proper annotations to describe the verified properties of the warnings and then prove them by  deductive verification. These approaches can precisely discard a number of FPs. However, not all of the generated warnings can be formally proved to be FPs or TPs by these approaches. Additional, model checking approaches also suffer from the enormous states space, which affect their performance and lead them to be non-scalable. 

In addition, several studies applied ML models to address SA warnings. Specially, some research~\cite{yuksel2014trust, flynn2018prioritizing, berman2019active} propose sets of features about statistic information of the warnings and then build a model which learn these features to classify SA warnings. 
However, these features are manually defined based on the dataset and the used SA tools. This process is error-prone even for experts. 
%
Meanwhile, instead of using a fixed set of features, Lee et al.~\cite{lee2019classifying} trained a CNN model classifying warnings based on features which are learned from lexical patterns in source code. 
However, they manually defined different contexts for different kinds of warnings based on their dataset. This limits the adaptation of their approach for other kinds of vulnerabilities and different dataset.
In this paper, we propose an approach which can be fully automated and easily to adapt for handling different warnings in different projects. Specially, our models are trained to capture the patterns associated with the warnings in their corresponding contexts, which are extracted by inter-procedural slicing techniques.

Moreover, ML are also actively adopted in vulnerabilities detection. Particularly, to leverage the  syntax and semantics information presented in the Abstract Syntax Tree, Dam et al.~\cite{dam2019lessons} proposed a deep learning tree based model to predict whether a source file is clean or defective. Besides, there are multiple studies also propose token-based models~\cite{russell2018automated, li2021sysevr} or graph-based models~\cite{zhou2019devign, chakraborty2021deep}  to predict whether a function containing vulnerabilities.  However, these research focuses on detecting vulnerabilities at the file level or function level, which are quite coarse-grained in granularity. Developers still need to investigate the whole source code in the detected files or functions to localize the vulnerabilities. 
In this research, our objective is more fine-grained in granularity. We focus on ranking the warnings which are reported by SA tools. With the resulting lists, developers can decide which vulnerabilities should be investigated and fixed in a given a amount of time.

\section{Conclusion}
\label{sec:conclusion}

SA tools have demonstrated their usefulness in detecting potential vulnerabilities. However, these tools often report a large number of warnings containing both TPs and FPs, which causes time-consuming for post-handling warnings and affects the productivity of developers.  
In this paper, we introduce \tool, a novel method for ranking SA warnings. Based on the reported statements and the corresponding warning contexts, we train two BiLSTM models  to capture the patterns associated with the TPs and FPs. After that, for a set of new warnings, \tool can predict the likelihood to be TP of each warning and then rank them according to the predicted scores. By using \tool, more actual vulnerabilities can be found in a given time. 
In order to evaluate the effectiveness of \tool, we conducted experiments on 6,620 warnings in 10 real-world projects. Our experimental results show that using \tool, developers can find +90\% of actual vulnerabilities by investigating only 60\% of the total warnings.

\section*{Acknowledgment}
This work has been supported by VNU University of Engineering and Technology under project number CN20.26.

In this work, \textit{Kien-Tuan Ngo} was funded by Vingroup Joint Stock Company and supported by the Domestic Master/ PhD Scholarship Programme of Vingroup Innovation Foundation (VINIF), Vingroup Big Data Institute (VINBIGDATA), code VINIF.2020.ThS.04.

\bibliographystyle{IEEEtran}
\bibliography{9.references}

\begin{thebibliography}{10}
\providecommand{\url}[1]{#1}
\csname url@samestyle\endcsname
\providecommand{\newblock}{\relax}
\providecommand{\bibinfo}[2]{#2}
\providecommand{\BIBentrySTDinterwordspacing}{\spaceskip=0pt\relax}
\providecommand{\BIBentryALTinterwordstretchfactor}{4}
\providecommand{\BIBentryALTinterwordspacing}{\spaceskip=\fontdimen2\font plus
\BIBentryALTinterwordstretchfactor\fontdimen3\font minus
  \fontdimen4\font\relax}
\providecommand{\BIBforeignlanguage}[2]{{%
\expandafter\ifx\csname l@#1\endcsname\relax
\typeout{** WARNING: IEEEtran.bst: No hyphenation pattern has been}%
\typeout{** loaded for the language `#1'. Using the pattern for}%
\typeout{** the default language instead.}%
\else
\language=\csname l@#1\endcsname
\fi
#2}}
\providecommand{\BIBdecl}{\relax}
\BIBdecl

\bibitem{ayewah2008using}
N.~Ayewah, W.~Pugh, D.~Hovemeyer, J.~D. Morgenthaler, and J.~Penix, ``Using
  static analysis to find bugs,'' \emph{IEEE software}, vol.~25, no.~5, pp.
  22--29, 2008.

\bibitem{nagappan2005static}
N.~Nagappan and T.~Ball, ``Static analysis tools as early indicators of
  pre-release defect density,'' in \emph{Proceedings. 27th International
  Conference on Software Engineering, 2005. ICSE 2005.}\hskip 1em plus 0.5em
  minus 0.4em\relax IEEE, 2005, pp. 580--586.

\bibitem{CERT}
\BIBentryALTinterwordspacing
C.~S.~C. Group, ``{SEI CERT Coding Standards (wiki)}.'' [Online]. Available:
  \url{https://wiki.sei.cmu.edu/confluence/display/seccode}
\BIBentrySTDinterwordspacing

\bibitem{MISRA}
M.~The Motor Industry Software Reliability~Association, \emph{Guidelines for
  the Use of the C Language in Critical Systems}, 03 2012.

\bibitem{beller2016analyzing}
M.~Beller, R.~Bholanath, S.~McIntosh, and A.~Zaidman, ``Analyzing the state of
  static analysis: A large-scale evaluation in open source software,'' in
  \emph{2016 IEEE 23rd International Conference on Software Analysis,
  Evolution, and Reengineering (SANER)}, vol.~1.\hskip 1em plus 0.5em minus
  0.4em\relax IEEE, 2016, pp. 470--481.

\bibitem{flynn2018prioritizing}
L.~Flynn, W.~Snavely, D.~Svoboda, N.~VanHoudnos, R.~Qin, J.~Burns, D.~Zubrow,
  R.~Stoddard, and G.~Marce-Santurio, ``Prioritizing alerts from multiple
  static analysis tools, using classification models,'' in \emph{2018 IEEE/ACM
  1st International Workshop on Software Qualities and their Dependencies
  (SQUADE)}.\hskip 1em plus 0.5em minus 0.4em\relax IEEE, 2018, pp. 13--20.

\bibitem{johnson2013don}
B.~Johnson, Y.~Song, E.~Murphy-Hill, and R.~Bowdidge, ``Why don't software
  developers use static analysis tools to find bugs?'' in \emph{2013 35th
  International Conference on Software Engineering (ICSE)}.\hskip 1em plus
  0.5em minus 0.4em\relax IEEE, 2013, pp. 672--681.

\bibitem{koc2019empirical}
U.~Koc, S.~Wei, J.~S. Foster, M.~Carpuat, and A.~A. Porter, ``An empirical
  assessment of machine learning approaches for triaging reports of a java
  static analysis tool,'' in \emph{2019 12th IEEE Conference on Software
  Testing, Validation and Verification (ICST)}.\hskip 1em plus 0.5em minus
  0.4em\relax IEEE, 2019, pp. 288--299.

\bibitem{ruthruff2008predicting}
J.~Ruthruff, J.~Penix, J.~Morgenthaler, S.~Elbaum, and G.~Rothermel,
  ``Predicting accurate and actionable static analysis warnings,'' in
  \emph{2008 ACM/IEEE 30th International Conference on Software
  Engineering}.\hskip 1em plus 0.5em minus 0.4em\relax IEEE, 2008, pp.
  341--350.

\bibitem{post2008reducing}
H.~Post, C.~Sinz, A.~Kaiser, and T.~Gorges, ``Reducing false positives by
  combining abstract interpretation and bounded model checking,'' in \emph{2008
  23rd IEEE/ACM International Conference on Automated Software
  Engineering}.\hskip 1em plus 0.5em minus 0.4em\relax IEEE, 2008, pp.
  188--197.

\bibitem{li2013software}
H.~Li, T.~Kim, M.~Bat-Erdene, and H.~Lee, ``Software vulnerability detection
  using backward trace analysis and symbolic execution,'' in \emph{2013
  International Conference on Availability, Reliability and Security}.\hskip
  1em plus 0.5em minus 0.4em\relax IEEE, 2013, pp. 446--454.

\bibitem{nguyen2019reducing}
T.~T. Nguyen, P.~Maleehuan, T.~Aoki, T.~Tomita, and I.~Yamada, ``Reducing false
  positives of static analysis for sei cert c coding standard,'' in \emph{2019
  IEEE/ACM Joint 7th International Workshop on Conducting Empirical Studies in
  Industry (CESI) and 6th International Workshop on Software Engineering
  Research and Industrial Practice (SER\&IP)}.\hskip 1em plus 0.5em minus
  0.4em\relax IEEE, 2019, pp. 41--48.

\bibitem{muske2016survey}
T.~Muske and A.~Serebrenik, ``Survey of approaches for handling static analysis
  alarms,'' in \emph{2016 IEEE 16th International Working Conference on Source
  Code Analysis and Manipulation (SCAM)}.\hskip 1em plus 0.5em minus
  0.4em\relax IEEE, 2016, pp. 157--166.

\bibitem{yuksel2014trust}
U.~Y{\"u}ksel, H.~S{\"o}zer, and M.~{\c{S}}ensoy, ``Trust-based fusion of
  classifiers for static code analysis,'' in \emph{17th International
  Conference on Information Fusion (FUSION)}.\hskip 1em plus 0.5em minus
  0.4em\relax IEEE, 2014, pp. 1--6.

\bibitem{berman2019active}
M.~Berman, S.~Adams, T.~Sherburne, C.~Fleming, and P.~Beling, ``Active learning
  to improve static analysis,'' in \emph{2019 18th IEEE International
  Conference On Machine Learning And Applications (ICMLA)}.\hskip 1em plus
  0.5em minus 0.4em\relax IEEE, 2019, pp. 1322--1327.

\bibitem{lee2019classifying}
S.~Lee, S.~Hong, J.~Yi, T.~Kim, C.-J. Kim, and S.~Yoo, ``Classifying false
  positive static checker alarms in continuous integration using convolutional
  neural networks,'' in \emph{2019 12th IEEE Conference on Software Testing,
  Validation and Verification (ICST)}.\hskip 1em plus 0.5em minus 0.4em\relax
  IEEE, 2019, pp. 391--401.

\bibitem{Mikolov2013EfficientEO}
T.~Mikolov, K.~Chen, G.~Corrado, and J.~Dean, ``Efficient estimation of word
  representations in vector space,'' in \emph{ICLR}, 2013.

\bibitem{okun2013report}
V.~Okun, A.~Delaitre, P.~E. Black \emph{et~al.}, ``Report on the static
  analysis tool exposition (sate) iv,'' \emph{NIST Special Publication}, vol.
  500, p. 297, 2013.

\bibitem{SARD}
\BIBentryALTinterwordspacing
N.~I. of~Standards and Technology, ``{Software assurance reference dataset}.''
  [Online]. Available: \url{https://samate.nist.gov/SRD/index.php}
\BIBentrySTDinterwordspacing

\bibitem{chakraborty2021deep}
S.~Chakraborty, R.~Krishna, Y.~Ding, and B.~Ray, ``Deep learning based
  vulnerability detection: Are we there yet,'' \emph{IEEE Transactions on
  Software Engineering}, 2021.

\bibitem{website}
\BIBentryALTinterwordspacing
``{DeFP}.'' [Online]. Available: \url{https://tuanngokien.github.io/DeFP/}
\BIBentrySTDinterwordspacing

\bibitem{FlawFinder}
\BIBentryALTinterwordspacing
``{Flawfinder}.'' [Online]. Available: \url{https://dwheeler.com/flawfinder/}
\BIBentrySTDinterwordspacing

\bibitem{horwitz1990interprocedural}
S.~Horwitz, T.~Reps, and D.~Binkley, ``Interprocedural slicing using dependence
  graphs,'' \emph{ACM Transactions on Programming Languages and Systems
  (TOPLAS)}, vol.~12, no.~1, pp. 26--60, 1990.

\bibitem{Joern}
\BIBentryALTinterwordspacing
``{Joern}.'' [Online]. Available: \url{https://docs.joern.io/home}
\BIBentrySTDinterwordspacing

\bibitem{lin2019software}
G.~Lin, J.~Zhang, W.~Luo, L.~Pan, O.~De~Vel, P.~Montague, and Y.~Xiang,
  ``Software vulnerability discovery via learning multi-domain knowledge
  bases,'' \emph{IEEE Transactions on Dependable and Secure Computing}, 2019.

\bibitem{hochreiter1997long}
S.~Hochreiter and J.~Schmidhuber, ``Long short-term memory,'' \emph{Neural
  computation}, vol.~9, no.~8, pp. 1735--1780, 1997.

\bibitem{zhou2019devign}
Y.~Zhou, S.~Liu, J.~Siow, X.~Du, and Y.~Liu, ``Devign: Effective vulnerability
  identification by learning comprehensive program semantics via graph neural
  networks,'' \emph{arXiv preprint arXiv:1909.03496}, 2019.

\bibitem{lin2019deep}
G.~Lin, W.~Xiao, J.~Zhang, and Y.~Xiang, ``Deep learning-based vulnerable
  function detection: A benchmark,'' in \emph{International Conference on
  Information and Communications Security}.\hskip 1em plus 0.5em minus
  0.4em\relax Springer, 2019, pp. 219--232.

\bibitem{CppCheck}
\BIBentryALTinterwordspacing
``{CppCheck}.'' [Online]. Available: \url{http://cppcheck.sourceforge.net}
\BIBentrySTDinterwordspacing

\bibitem{Rats}
\BIBentryALTinterwordspacing
``{RATS - Rough Auditing Tool for Security}.'' [Online]. Available:
  \url{https://github.com/andrew-d/rough-auditing-tool-for-security}
\BIBentrySTDinterwordspacing

\bibitem{dam2019lessons}
H.~K. Dam, T.~Pham, S.~W. Ng, T.~Tran, J.~Grundy, A.~Ghose, T.~Kim, and C.-J.
  Kim, ``Lessons learned from using a deep tree-based model for software defect
  prediction in practice,'' in \emph{2019 IEEE/ACM 16th International
  Conference on Mining Software Repositories (MSR)}.\hskip 1em plus 0.5em minus
  0.4em\relax IEEE, 2019, pp. 46--57.

\bibitem{nguyen2021comparison}
H.~N. Nguyen, S.~Teerakanok, A.~Inomata, and T.~Uehara, ``The comparison of
  word embedding techniques in rnns for vulnerability detection.'' in
  \emph{International Conference on Information Systems Security and Privacy
  (ICISSP)}, 2021, pp. 109--120.

\bibitem{li2018vuldeepecker}
Z.~Li, D.~Zou, S.~Xu, X.~Ou, H.~Jin, S.~Wang, Z.~Deng, and Y.~Zhong,
  ``Vuldeepecker: A deep learning-based system for vulnerability detection,''
  \emph{arXiv preprint arXiv:1801.01681}, 2018.

\bibitem{baca2009static}
D.~Baca, K.~Petersen, B.~Carlsson, and L.~Lundberg, ``Static code analysis to
  detect software security vulnerabilities-does experience matter?'' in
  \emph{2009 International Conference on Availability, Reliability and
  Security}.\hskip 1em plus 0.5em minus 0.4em\relax IEEE, 2009, pp. 804--810.

\bibitem{muske2013efficient}
T.~Muske, A.~Datar, M.~Khanzode, and K.~Madhukar, ``Efficient elimination of
  false positives using bounded model checking,'' in \emph{ISSRE}, vol.~15,
  2013, pp. 2--5.

\bibitem{muske2015efficient}
T.~Muske and U.~P. Khedker, ``Efficient elimination of false positives using
  static analysis,'' in \emph{2015 IEEE 26th International Symposium on
  Software Reliability Engineering (ISSRE)}.\hskip 1em plus 0.5em minus
  0.4em\relax IEEE, 2015, pp. 270--280.

\bibitem{russell2018automated}
R.~Russell, L.~Kim, L.~Hamilton, T.~Lazovich, J.~Harer, O.~Ozdemir,
  P.~Ellingwood, and M.~McConley, ``Automated vulnerability detection in source
  code using deep representation learning,'' in \emph{2018 17th IEEE
  international conference on machine learning and applications (ICMLA)}.\hskip
  1em plus 0.5em minus 0.4em\relax IEEE, 2018, pp. 757--762.

\bibitem{li2021sysevr}
Z.~Li, D.~Zou, S.~Xu, H.~Jin, Y.~Zhu, and Z.~Chen, ``Sysevr: A framework for
  using deep learning to detect software vulnerabilities,'' \emph{IEEE
  Transactions on Dependable and Secure Computing}, 2021.

\end{thebibliography}
\end{document}